\documentclass[a4paper,12pt]{article}
\usepackage{graphicx,amsmath,amssymb,bm,latexsym,mathrsfs,amsfonts}
\newcommand{\bea}{\begin{eqnarray}}
\newcommand{\ena}{\end{eqnarray}}
\newcommand{\vs}[1]{\vspace{#1 mm}}

\renewcommand{\a}{\alpha}
\renewcommand{\b}{\beta}
\renewcommand{\c}{\gamma}
\renewcommand{\d}{\delta}

\newcommand{\dsl}{\pa \kern-0.5em /}

\newcommand{\pa}{\partial}

\newcommand{\nn}{\nonumber\\}
\newcommand{\p}[1]{(\ref{#1})}
\newcommand{\lan}{\langle}
\newcommand{\ran}{\rangle}
\newcommand{\st}{\stackrel}
\newcommand{\xx}{\!\!\!\!\!\!\!\!\!\!\!}
\newcommand{\x}{\!\!\!\!\!\!}

\begin{document}

\begin{titlepage}

\begin{flushright}
KU-TP 054
\end{flushright}

\vs{10}
\begin{center}
{\large\bf Hawking Radiation and Tunneling Mechanism
for a New Class of Black Holes in Einstein-Gauss-Bonnet Gravity}
\vs{15}

{\large
Kenji Muneyuki
and
Nobuyoshi Ohta\footnote{e-mail address: ohtan@phys.kindai.ac.jp}
} \\
\vs{10}
{\em Department of Physics, Kinki University,
Higashi-Osaka, Osaka 577-8502, Japan}

\vs{15}
{\bf Abstract}
\end{center}
\vs{5}

We study the Hawking radiation in a new class of black hole solutions
in the Einstein-Gauss-Bonnet theory. The black hole has been argued to have
vanishing mass and entropy, but finite Hawking temperature.
To check if it really emits radiation, we analyse the Hawking radiation using
the original method of quantization of scalar field in the black hole background and
the quantum tunneling method, and confirm that it emits radiation at the Hawking
temperature. A general formula is derived for the Hawking temperature and backreaction
in the tunneling approach.
Physical implications of these results are discussed.

\end{titlepage}
\newpage
\renewcommand{\thefootnote}{\arabic{footnote}}
\setcounter{page}{2}

\newpage

\section{Introduction}
\label{Introduction}

Hawking radiation from the black holes is one of the most striking effects to arise
from the combination of quantum mechanics and general relativity.
In its original derivation~\cite{H}, Hawking considered quantization
of matter in a background spacetime that contains an event horizon and
found that the occupation number spectrum of quantum modes in the vacuum state
is that of a blackbody at a fixed temperature, now called Hawking temperature.
Since then there appear several derivations, including those based on
quantum tunneling~\cite{PW} and gravitational anomalies~\cite{RW}.
These effects are very important in revealing quantum aspects of gravitational theories.

Recently a new class of black hole solutions in the Einstein-Gauss-Bonnet gravity has been
found in \cite{MD}, and their higher-dimensional generalization is considered in \cite{CCO}.
The solution consists of spacetime with the topology of the direct product of
the $(n-4)$-dimensional space of constant curvature $\mathcal{K}^{n-4}$ and
the usual four-dimensional spacetime $\mathcal{M}^4$.
The vacuum equations with Gauss-Bonnet contribution and cosmological constant $\Lambda$
are split into four-dimensional part and extra $(n-4)$-dimensional part.
The four-dimensional part is like the vacuum Einstein equation with a cosmological
constant redefined while the extra dimensional part gives a scalar constraint equation.
The solution represents a black hole with two horizons and $\mathcal{M}^4$
asymptotically approaches Reissner-Nortstr\"{o}m-(Anti)-de Sitter (RN-(A)dS)
spacetime or RN spacetime for positive Gauss-Bonnet coupling constant.
We can also choose parameters so as to obtain asymptotically flat solutions.
One very curious property of this black hole is that it has zero mass and entropy,
but the Hawking temperature is nonzero. As argued in Ref.~\cite{CCO}, the former
property follows from the higher-dimensional nature of the black hole.
However it may seem strange if the black hole emits any radiation. If it has no energy,
it is natural to expect that there is no radiation and the temperature is zero.
Note that the first law of thermodynamics is satisfied, so there is no inconsistency.
In \cite{CCO}, the Hawking temperature was evaluated by making the analytic continuation to
the Euclidean geometry and read from the periodicity in the Euclidean time around
the horizon~\cite{GH}.
The natural question then arises whether this method is still valid in this new black
hole solution and there is really radiation since its presence is not explicitly shown.
Thus it is interesting and important to study if the black hole really
emits radiation or not and examine quantum aspects of the solution.
In this paper we investigate this problem using the original method of Hawking and
another method based on tunneling mechanism.

The derivation of Hawking radiation based on the quantization of matter fields gives
rather intuitive picture, while the one using tunneling mechanism has the advantage
that it can easily give backreaction. There is a vast literature on this method.
We list some of works relevant to our later discussions~\cite{HL1}-\cite{VAD}.
The method considers that a  pair of particle and anti-particle is formed close to the horizon
inside a black hole. The particle of outgoing modes tries to move outside the black hole,
while the anti-particle of ingoing modes moves toward the center of the black hole.
The horizon plays a role of a barrier when the particle tries to move outside
the black hole, but the particle can get out of the black hole by
the quantum tunneling effect. The WKB probability amplitude for the particle
is calculated by taking into account classically forbidden paths. By comparing with
the Boltzmann factor in a thermal equilibrium state, the Hawking temperature is obtained.
To check if there is really radiation, we derive Hawking radiation for the new
black holes mentioned above both in the original and tunneling approaches,
and compare the results. We find that both approaches confirm that the black hole really
emits radiation at a certain temperature. This is somewhat surprising result.

The paper is organized as follows. In Sect.~\ref{Schwarzschild},
we briefly review the Schwarzschild-like solution for $d=4$ and $n\geq8$ in the
Einstein-Gauss-Bonnet theory and give the argument why it has zero mass and entropy
but finite Hawking temperature. In Sect.~\ref{Hrad}, we calculate Hawking radiation
by using two methods. We also give an alternative general
formula which simplifies the calculation and clarifies the connection
between the Hawking temperature and the surface gravity.
Finally we discuss these somewhat surprising results
and draw our conclusions in Sect.~\ref{Conclusion}.

\section{Schwarzschild-like solutions in the Einstein-Gauss-Bonnet gravity}
\label{Schwarzschild}

We consider the action for $n$-dimensional spacetime~\cite{MD}:
\begin{align}
S=\frac{1}{16\pi G_{n}}\int d^n x \sqrt{-g}\left(R-2\Lambda+\a L_{GB}\right),
\label{2-1}
\end{align}
where $G_{n}$ is the $n$-dimensional gravitational constant, $\a$ the Gauss-Bonnet
coupling constant, $R$ the $n$-dimensional Ricci scalar and $\Lambda$ the cosmological
constant. The Gauss-Bonnet term $L_{GB}$ is given by
\begin{align}
L_{GB}\equiv R^2-4R_{\mu\nu}R^{\mu\nu}+R_{\mu\nu\rho\sigma}R^{\mu\nu\rho\sigma}
.
\label{2-2}
\end{align}
We take $\a\geq0$, because it is so in superstring models.

The gravitational equation from the action (\ref{2-1}) is given by
\begin{align}
\mathcal{G}^{\mu}{}_\nu
\equiv G^\mu{}_\nu+\a H^\mu{}_\nu+\Lambda\d^\mu{}_\nu
=0,
\label{2-3}
\end{align}
where
\bea
G_{\mu\nu} &\equiv& R_{\mu\nu}-\frac{1}{2}g_{\mu\nu}R, \nn
H_{\mu\nu} &\equiv& 2\left(RR_{\mu\nu}-2R_{\mu\a}R^\a{}_\nu-2R^{\a\b}R_{\mu\a\nu\b}
+R_\mu{}^{\a\b\c}R_{\nu\a\b\c}\right)-\frac{1}{2}g_{\mu\nu}L_{GB}.
\label{2-5}
\ena

We consider the solutions to the above $n$-dimensional vacuum gravitational equation,
which are locally homeomorphic to $\mathcal{M}^4\times\mathcal{K}^{n-4}$ with
the metric $g_{\mu\nu}=\mathrm{diag}(g_{AB},r_{0}^2\c_{ab})$.
Here $g_{AB}\;(A,B=0,\cdots,3)$ is an arbitrary Lorentz
metric on $\mathcal{M}^4$, $r_{0}$ is a constant and $\c_{ab}\;(a,b=4,\cdots,n-1)$
is the unit metric on the $(n-4)$-dimensional Einstein space $\mathcal{K}^{n-4}$
with constant curvature $\bar{k}=-1$.
Then the vacuum gravitational equation $\mathcal{G}^A{}_B=0$ is a tensorial
equation on $\mathcal{M}^4$, while $\mathcal{G}^a{}_b/\d^a{}_b=0$
is a scalar equation on $\mathcal{M}^{n-4}$.

To solve these equations, we assume the Einstein-space condition~\cite{MD}:
\begin{align}
\st{\xx\x(n-4)}{C^{\;abcd}}\quad\st{\xx\x(n-4)}{C_{\;fbcd}}=\Theta \d^a_{\;\; f},
\label{2-9}
\end{align}
where $\st{\x\x(n-4)}{C^{\;abcd}}$ is the $(n-4)$-dimensional Weyl tensor
in $\mathcal{K}^{n-4}$.
$\Theta$ is a constant by the consistency with the identity $H^{\mu\nu}{}_{;\nu}=0$.
The Weyl tensor vanishes identically in three or less dimension, so that
$\Theta\equiv0$ holds for $n\leq 7$.

In this paper, we consider $n\geq8$ and impose the conditions
(i) $r_{0}^2= 2\a(n-4)(n-5)$ and
(ii) $\a\Lambda=-\frac{n^2-5n-2}{8(n-4)(n-5)}+\frac{\Theta}{8(n-4)(n-5)^2}$.
With conditions (i) and (ii), the four-dimensional part of the vacuum gravitational
equation $\mathcal{G}^A{}_B=0$ is automatically satisfied, and the $(n-4)$-dimensional part
$\mathcal{G}^a{}_b=0$ gives a single scalar equation on $\mathcal{M}^4$,
which determines the metric on $\mathcal{M}^4$.
The constant $r_{0}^2$ is proportional to the Gauss-Bonnet coupling
constant $\a$, which is supposed to be of the order of the square of the Planck
length. Thus, compactifying $\mathcal{K}^{n-4}$ by appropriate identifications,
we obtain the Kaluza-Klein spacetime with small and compact extra dimensions.

The metric on $\mathcal{M}^4$ for the static solution is given by
\begin{align}
ds^2=g_{AB}dx^Adx^B=-f(r)dt^2+\frac{1}{f(r)}dr^2+r^2d\Sigma^2_{2(k)},
\label{3-1}
\end{align}
where
\begin{align}
\label{3-2}
f(r)&=k+\frac{r^2}{2(n-4)\a}\left[1\mp\left\{1-\frac{2(n-5)(2n-11)-\Theta}{6(n-5)^2}
-\frac{4(n-4)\a m}{r^3}-\frac{4(n-4)\a q}{r^4}\right\}^{\frac{1}{2}}\right],\\
& d\Sigma^2_{2(k)}=
\left\{
\begin{array}{lll}
d\theta^2+\sin^2\theta d\phi^2 & \mbox{for} &k=1 \\
d\theta^2+ d\phi^2 & \mbox{for} & k=0 \\
d\theta^2+\sinh^2\theta d\phi^2 & \mbox{for} &k=-1
\end{array}
\right.
\end{align}
where $m$ and $q$ are integration constants, and $d\Sigma^2_{2(k)}$
is the line element on the unit Einstein space $\mathcal{K}^2$ and $k=\pm1,0$.
There are two branches of the solution
corresponding to the sign in front of the square root in (\ref{3-2}),
which we call the minus- and plus-branches.

The metric on $\mathcal{M}^4$ is asymptotically RN-(A)dS
spacetime for $k=1$ in spite of absence of Maxwell field since the function
$f(r)$ behaves for $r\rightarrow\infty$ as
\begin{align}
f(r)\simeq\frac{r^2}{2(n-4)\a}\Bigg(1\mp\sqrt{\frac{2(n-4)(n-5)+\Theta}{6(n-5)^2}}\Bigg)
+k\mp\sqrt{\frac{6(n-5)^2}{2(n-4)(n-5)+\Theta}}\Bigg(-\frac{m}{r}-\frac{q}{r^2}\Bigg).
\label{3-3}
\end{align}
Here we are interested in the asymptotically flat spacetime.
This can be obtained by taking the minus-branch, and setting $k=1$, $\Theta=2(n-5)(2n-11)$
and $m=-2G_{4}M$ in (\ref{3-2}):
\bea
f(r)&=&1+\frac{r^2}{2(n-4)\a}\left[1-\sqrt{1+\frac{8(n-4)\a G_{4}M}{r^3}
-\frac{4(n-4)\a q}{r^4}}\,\right] \nn
&\sim & 1-\frac{2 G_4 M}{r}-\frac{q}{r^2}.
\label{4-2}
\ena
Note that even though we have a cosmological constant, the flat space is obtained due
to the presence of the higher derivative terms.

By examining the asymptotic behavior of this function, one might conclude that
the black hole has a mass proportional to $M$.
However, this black hole solution is quite different from the one obtained
by the usual Kaluza-Klein dimensional reduction~\cite{CCO}. In the usual Kaluza-Klein theory
with a direct product manifold ${\cal M}^4 \times {\cal K}^{n-4}$, we get reduced
action by integrating the total action over the extra space ${\cal K}^{n-4}$,
and then derive field equation in four dimensions whose solution gives the black hole.
However, the above solution trivially satisfies this four-dimensional part
of the field equation, and the nontrivial part is the trace part for the $(n-4)$ dimensions,
which cannot be obtained from the reduced action.
Thus the reduced action does not give any information on the above solution.
The usual way to define the mass in four dimensions using
the corresponding four-dimensional locally diffeomorphism-invariant
``effective action'' does not work, and we have to use the whole $n$-dimensional theory
to derive the mass and entropy. Namely we cannot simply read off
the mass from the asymptotic behavior of the four-dimensional solution~\p{4-2},
and the entropy should not be evaluated only by using the action in four dimensions.
Instead we should use Wald's entropy formula in the whole dimensions~\cite{Wald}
\bea
S = \frac{1}{4 G_n} \int d^{n-2}x \sqrt{-g} (1 + 2 \a R)(\tilde h),
\label{ent}
\ena
where the metric is replaced by $\tilde h$, which is the induced metric on
the $(n-2)$-dimensional cross section of the horizon. The curvature is found to be
\bea
R( \tilde h) = \overset{(4)}{R}-(n-4)(n-5)\frac{1}{r_0^2},
\label{curvdec}
\ena
where $\overset{(4)}{R}$ is the four-dimensional scalar curvature computed
using the metric~\p{3-1}. Using the explicit value of $r_0$, we find that
the second term in \p{curvdec} cancels against the first term 1 in the entropy~\p{ent},
and the remaining term gives a constant (proportional to the Euler number)
independent of the horizon radius. This constant should be
discarded~\cite{CCO}. There are several reasons for this; obvious one is that
such a constant remains even in the vanishing limit of the horizon radius
which should give zero. On the other hand, we get nonzero Hawking temperature
if we evaluate it by Euclideanization method.
Thus, surprisingly enough, it is found that
the entropy vanishes for the solution, but the Hawking temperature,
evaluated by the Euclideanization technique, is nonvanishing.

Similarly, using the Euclidean action of higher dimensions, we can evaluate
the mass of the black hole by evaluating the Euclidean action, which gives
a constant independent of the temperature.  Considering the relation between the
Euclidean action and free energy, we immediately see that the energy of the black
hole vanishes~\cite{CCO}.
These results, vanishing mass $M$ and entropy $S$ but nonzero Hawking temperature $T_H$,
are obviously consistent with the first law of thermodynamics $dM=T_H dS$.
This raises an important question whether the Hawking radiation really takes place or not
even though the temperature appears finite since its presence is not explicitly confirmed.
This is what we now study.

If one naively evaluated the entropy in the four-dimensional viewpoint using the first
law of thermodynamics, one would find that the entropy would be given by an expression
containing logarithm of the horizon area. However, as we argued above, this method is
based on the effective four-dimensional action which does not capture the properties
of our black holes, and cannot be justified. For more details, we refer the reader
to Ref.~\cite{CCO}.

\section{Hawking radiation and useful formula}
\label{Hrad}

\subsection{Particle creation}
\label{Particle creation}

Let us first calculate the Hawking radiation for the metric on $\mathcal{M}^4$
which is asymptotically RN spacetime without Maxwell field
and show that there exists the Hawking radiation in four dimensions with a temperature
given by the surface gravity, just as the usual Schwarzschild solution~\cite{H}.

The Klein-Gordon equation in a curved spacetime is given by
\begin{align}
\frac{1}{\sqrt{-g}}\pa_A\left(g^{AB}\sqrt{-g}\pa_{B}\Phi\right)=0,
\label{4-3}
\end{align}
where $\Phi\equiv\Phi(t,r,\theta,\phi)$.
Inserting the metric (\ref{3-1}), we obtain the following differential equation
\begin{align}
\left[-\frac{1}{f(r)}\frac{\pa^2}{\pa t^2}+\frac{1}{r^2}\frac{\pa}{\pa r}
\left\{r^2f(r)\frac{\pa}{\pa r}\right\}+\frac{1}{r^2}\left\{\frac{1}{\sin\theta}
\frac{\pa}{\pa \theta}\left(\sin\theta\frac{\pa}{\pa \theta}\right)
+\frac{1}{\sin^2\theta}\frac{\pa^2}{\pa \phi^2}\right\}\right]\Phi=0
\label{4-4}
\end{align}
Using spherical symmetry and time translation invariance, we write the scalar field as
\begin{align}
\Phi(t,r,\theta,\phi)
=\left(Ae^{-i\omega t}+A^{\ast}e^{i\omega t}\right)R(r)Y_{lm}(\theta,\phi),
\label{4-5}
\end{align}
where $A$ is constant and $Y_{lm}(\theta,\phi)$ is the spherical harmonic function.
We do not have to consider the angular part of the solution because the spacetime
$\mathcal{M}^{4}$ is independent of $\theta$ and $\phi$.
We insert (\ref{4-5}) in (\ref{4-4}) and introduce the new field $R(r)\equiv \tilde{R}(r)/r$
to obtain
\bea
\frac{\pa^2 \tilde{R}(r)}{\pa r^{\ast^2}}+\omega^2\tilde{R}(r)-f(r)
\left[\frac{1}{r}\frac{\pa f(r)}{\pa r}+\frac{l(l+1)}{r^2}\right]\tilde{R}(r)=0,
\label{4-9}
\ena
where
\bea
\frac{\pa f(r)}{\pa r} &=& \frac{1}{2(n-4)\a}\left[2r-\frac{2r^3+4(n-4)\a M}{\sqrt{r^4
+8(n-4)\a Mr-4(n-4)\a q}}\right],
\label{4-10}
\\
r^{\ast} &\equiv& \int\frac{1}{f(r)}dr.
\label{4-8}
\ena
In the limit of $r\rightarrow \infty$, Eq.~\p{4-9} reduces to
\begin{align}
\frac{\pa^2 \tilde{R}(r)}{\pa r^{\ast^2}}+\omega^2\tilde{R}(r)=0.
\label{4-11}
\end{align}
Thus, the field (\ref{4-5}) can be expanded in the initial stationary region as
\begin{align}
\Phi=\sum_{\omega} (a_{\omega}^{in}f_{\omega}+a_{\omega}^{in\dagger}f_{\omega}^{\ast}),
\label{4-12}
\end{align}
or in the final one as
\begin{align}
\Phi=\sum_{\omega} (a_{\omega}^{out}p_{\omega}+a_{\omega}^{out\dagger}p_{\omega}^{\ast}),
\label{4-13}
\end{align}
where $a_{\omega}^{in}$ and $a_{\omega}^{out}$ are the annihilation operators
which satisfy the usual commutation relations, and $f_{\omega}$ and $p_{\omega}$
are the solutions of (\ref{4-11}) in the initial and final regions, respectively.
The positive frequency modes for (\ref{4-11}) at past null infinity ($I^{-}$) are
\begin{align}
f_{\omega}(r,v)=\frac{e^{-i\omega(t+r^{\ast})}}{4\pi r\sqrt{\omega}}
=\frac{e^{-i\omega v}}{4\pi r\sqrt{\omega}},
\label{4-14}
\end{align}
where $v\equiv t+r^{\ast}$ and they have the scalar product
\begin{align}
(f_{\omega},f_{\omega^{\prime}})
&=-(f_{\omega}^{\ast},f_{\omega^{\prime}}^{\ast})
=-i\int_{I^{-}}dvr^2d\Sigma_{k=1}(f_{\omega}\pa_{v}f_{\omega^{\prime}}^{\ast}
-f_{\omega^{\prime}}^{\ast}\pa_{v}f_{\omega})\nn
&=\d (\omega-\omega^{\prime}),
\label{4-15}
\end{align}
and $(f_{\omega},f_{\omega^{\prime}}^{\ast})=0$.
Similarly the positive frequency modes for (\ref{4-11}) at future null infinity ($I^{+}$) are
\begin{align}
p_{\omega}(r,u)=\frac{e^{-i\omega(t-r^{\ast})}}{4\pi r\sqrt{\omega}}
=\frac{e^{-i\omega u}}{4\pi r\sqrt{\omega}},
\label{4-16}
\end{align}
where $u\equiv t-r^{\ast}$ and they have the scalar product
\begin{align}
(p_{\omega},p_{\omega^{\prime}})
&=-(p_{\omega}^{\ast},p_{\omega^{\prime}}^{\ast})
=-i\int_{I^{+}}dvr^2d\Sigma_{k=1}(p_{\omega}\pa_{v}p_{\omega^{\prime}}^{\ast}
-p_{\omega^{\prime}}^{\ast}\pa_{v}p_{\omega})\nn
&=\d (\omega-\omega^{\prime}),
\label{4-17}
\end{align}
and $(p_{\omega},p_{\omega^{\prime}}^{\ast})=0$.

We can now calculate the Bogoliubov coefficients relating the ingoing and outgoing
solutions $f_{\omega}$ and $p_{\omega}$:
\begin{align}
&p_{\omega}=\sum_{\omega^{\prime}} (A_{\omega\omega^{\prime}}f_{\omega^{\prime}}
+B_{\omega\omega^{\prime}}f_{\omega^{\prime}}^{\ast}),
\label{4-18}\\
&A_{\omega\omega^{\prime}}=(p_{\omega},f_{\omega^{\prime}})
\mbox{ and }
B_{\omega\omega^{\prime}}=-(p_{\omega},f_{\omega^{\prime}}^{\ast}),
\label{4-19}
\end{align}
The Bogoliubov coefficients $A$ and $B$ are used to expand one of the two sets of
creation and annihilation operators in terms of the other:
\begin{align}
&a_{\omega}^{in}=\sum_{\omega^{\prime}} (A_{\omega^{\prime}\omega}a_{\omega^{\prime}}^{out}
+B_{\omega^{\prime}\omega}a_{\omega^{\prime}}^{out\dagger}),
\label{4-21}\\
&a_{\omega}^{out}=\sum_{\omega^{\prime}} (A_{\omega\omega^{\prime}}^{\ast}
a_{\omega^{\prime}}^{in}-B_{\omega\omega^{\prime}}a_{\omega^{\prime}}^{in\dagger}).
\label{4-22}
\end{align}
Thus if $B_{\omega\omega^{\prime}}$ is non-zero, the particle content of the vacuum
state $\rvert\mathrm{vac}\ran_{in}$ at $I^{-}$ is nontrivial:
\begin{align}
_{in}\lan\mathrm{vac}\lvert N_{\omega}^{+}\rvert\mathrm{vac}\ran_{in}
=\sum_{\omega^{\prime}}\lvert B_{\omega\omega^{\prime}}\rvert^2,
\label{4-23}
\end{align}
where $N_{\omega}^{+}$ is the particle number operator at frequency $\omega$ at $I^{+}$.

In $I^{-}$, we find
\begin{align}
&p_{\omega}=\frac{e^{i\frac{\omega}{\kappa}\mbox{ln}(-v)}}{4\pi r\sqrt{\omega}}
&\mbox{for }v<0,\nn
&p_{\omega}=0&\mbox{for }v>0,
\label{4-24}
\end{align}
where we have used the surface gravity $\kappa$ on the horizon defined by
\begin{align}
\kappa\equiv\frac{1}{2}\pa_rf(r)\Bigr|_{r=r_{+}}=\frac{r_{+}-M}{r_{+}^2+2(n-4)\a}.
\label{4-25}
\end{align}
Inserting (\ref{4-14}) and (\ref{4-24}) in (\ref{4-19}), we obtain
\bea
\lvert A_{\omega\omega^{\prime}}\rvert
=e^{\frac{\pi\omega}{\kappa}}\lvert B_{\omega\omega^{\prime}}\rvert.
\label{4-26}
\ena
It follows from (\ref{4-26}) that
\begin{align}
\d_{\omega\omega^{\prime}}&=(AA^{\dagger})_{\omega\omega^{\prime}}
-(BB^{\dagger})_{\omega\omega^{\prime}}\nn
&=\left[e^{\frac{\pi(\omega+\omega^{\prime})}{\kappa}}-1\right]
(BB^{\dagger})_{\omega\omega^{\prime}}.
\label{4-27}
\end{align}
Setting $\omega=\omega^{\prime}$, we find
\begin{align}
_{in}\lan\mathrm{vac}\lvert N_{\omega}^{+}\rvert\mathrm{vac}\ran_{in}
=(BB^{\dagger})_{\omega\omega}=\frac{1}{e^{\frac{2\pi\omega}{\kappa}}-1}.
\label{4-28}
\end{align}
This is the Planck distribution for black hole radiation with the Hawking temperature
\begin{align}
T_{H}=\frac{\kappa}{2\pi}.
\label{4-29}
\end{align}
Thus we confirm that the radiation is emitted.
The temperature agrees with the one obtained by Euclideanization~\cite{CCO,GH}.

\subsection{Tunneling mechanism}
\label{Tunneling mechanism}

To check the validity of the above result further, let us next consider Hawking radiation
as a quantum tunneling process through the horizon, following Ref.~\cite{U,HL10}.
We should use a coordinate system that is not singular at the horizon,
because a particle passes through the horizon without singularity on the path.
The outer (inner) horizon of the black hole is defined by
\begin{align}
r_{\pm}=M\pm\sqrt{M^2-\left(q+(n-4)\a\right)}.
\label{5-3}
\end{align}
Painlev\'{e} coordinates which are used to eliminate coordinate singularity
are convenient choices in this analysis.
With the Painlev\'{e} time transformation
\begin{align}
dt\rightarrow dt-\frac{\sqrt{1-f(r)}}{f(r)}dr.
\label{5-4}
\end{align}
the Painlev\'{e}-like line element is given by
\begin{align}
ds^2=-f(r)dt^2+2\sqrt{1-f(r)}dtdr+dr^2+r^2d\Sigma^2_{2(k=1)}.
\label{5-5}
\end{align}
There is no coordinate singularity at the horizon.

The outgoing motion of the massless particle (the outgoing radial null geodesics
$ds^2=d\Sigma^2_{2(k=1)}=0$) takes the form
\begin{align}
\dot{r}\equiv \frac{dr}{dt}=1-\sqrt{1-f(r)}.
\label{5-6}
\end{align}
When we take into account the effect of the particle's self-gravitation,
we should replace $M$ by $M-\omega$, where $\omega$ is the energy of the particle
which escapes from the black hole by the tunneling mechanism.
Then (\ref{5-5}) and (\ref{5-6}) are rewritten as
\begin{align}
ds^2&=-f_a(r)dt^2+2\sqrt{1-f_a(r)}dtdr+dr^2+r^2d\Sigma^2_{2(k=1)},
\label{5-7}\\
\dot{r}&=1-\sqrt{1-f_a(r)},
\label{5-8}
\end{align}
where
\begin{align}
f_a(r)&=1+\frac{r^2}{2(n-4)\a}\left[1-\sqrt{1+\frac{8(n-4)\a (M-\omega)}{r^3}
-\frac{4(n-4)\a q}{r^4}}\right].
\label{5-9}
\end{align}

We evaluate the WKB probability amplitude for a classically forbidden trajectory.
The imaginary part of the action for an outgoing positive energy particle,
which crosses the horizon outwards from $r_{in}$ to $r_{out}$, is given by
\begin{align}
\mbox{Im }S =\mbox{Im}\int_{r_{in}}^{r_{out}}p_{r}dr
=\mbox{Im}\int_{r_{in}}^{r_{out}}\int_{0}^{p_{r}}dp_{r}^{\prime}dr.
\label{5-10}
\end{align}
Use of Hamilton's equation $\dot{r}=\frac{dH}{dp_{r}}\vert_r$ in (\ref{5-10}) yields
\begin{align}
\mbox{Im }S&=\mbox{Im}\int_{M}^{M-\omega}\int_{r_{in}}^{r_{out}}\frac{dr}{\dot{r}}dH
=\mbox{Im}\int_0^{\omega}\int_{r_{in}}^{r_{out}}\frac{dr}{1-\sqrt{1-f_a(r)}}
(-d\omega')\nn
&=- \mbox{Im}\int_0^{\omega}\int_{r_{in}}^{r_{out}}
\frac{g(M-\omega^{\prime})}{(M-\omega^{\prime})-\frac{1}{2r}[r^2+(q+(n-4)\a)]}
dr d\omega',
\label{5-11}
\end{align}
where
\begin{align}
& g(M-\omega^{\prime})
\equiv -\frac{\Big\{ 1+\left[\frac{1}{2(n-4)\a} \left( h^{1/2}- r^2\right)
\right]^{\frac{1}{2}} \Big\}
\left\{ r^2 + 2(n-4)\a + h^{1/2}\right\}}{4r},\nn
& h \equiv r^4+8(n-4)\a (M-\omega^{\prime})r-4(n-4)\a q,\nonumber
\end{align}
and use has been made of $H=M-\omega^{\prime}$.

Using Feynman's $i\epsilon$ prescription $\omega\rightarrow\omega-i\epsilon$
and performing $M$-integration, we find
\begin{align}
\mbox{Im }S
=-\pi\int_{r_{in}}^{r_{out}}\frac{r^2+2(n-4)\a}{r}dr,
\label{5-12}
\end{align}
where $r_{out}$ and $r_{in}$ are given by
\begin{align}
r_{out}&=M-\omega+\sqrt{(M-\omega)^2-\left(q+(n-4)\a\right)}\nn
&\simeq r_{+}-\frac{r_{+}}{r_{+}-M}\omega+\frac{r_{+}(r_{+}-2M)}{2(r_{+}-M)^3}\omega^2,
\label{5-13}\\
r_{in}&=r_{+}=M+\sqrt{M^2-\left(q+(n-4)\a\right)},
\label{5-14}
\end{align}
respectively. To the second order in $\omega$, Eq.~(\ref{5-12}) gives
\begin{align}
\mbox{Im }S&=-\frac{\pi}{2}(r_{out}^2-r_{in}^2)-2(n-4)\a
\ln\left(\frac{r_{out}}{r_{in}}\right)\nn
&\simeq\left(\frac{r_{+}^2+2(n-4)\a}{r_{+}-M}\right)\pi\omega
-\left(\frac{2r_{+}^3-3Mr_{+}^2-2(n-4)\a M}{2(r_{+}-M)^3}\right)\pi\omega^2.
\label{5-15}
\end{align}
Thus, we obtain the WKB probability amplitude
\begin{align}
\Gamma=e^{-2\, {\rm Im\, }S}\simeq e^{-\frac{2\pi}{\kappa}\omega
+\left(\frac{2r_{+}^3-3Mr_{+}^2-2(n-4)\a M}{(r_{+}-M)^3}\right)\pi\omega^2}.
\label{5-16}
\end{align}
{}From the first order in $\omega$ in (\ref{5-15}), by comparing the result~(\ref{5-15})
with the Boltzmann factor ($\Gamma=e^{-\frac{\omega}{T}}$) in a thermal equilibrium
state at temperature $T$, we find the Hawking temperature $T_{H}$ is given by
\begin{align}
T_{H}=\frac{\kappa}{2\pi}.
\label{5-18}
\end{align}
We thus find that the radiation exists and the result (\ref{5-18}) is consistent
with the result (\ref{4-29}).
The advantage of this method is that we can easily obtain the effect to the next order
in $\omega$.

\subsection{Relation to surface gravity and backreaction}

This second derivation of the Hawking temperature is a little involved.
Due to this, though we see that the result agrees with the first evaluation,
it is not so obvious whether it gives result consistent with the first one
for more general case.
Here we give a simpler evaluation of the amplitude which gives directly the
result as a surface gravity, so it becomes apparent to give the same result.
Not only that, we can also get a formula for the additional term higher in $\omega$.

When $r$ is close to the horizon and $M\gg\omega$, we can use the fact that $f$ has
a zero at the horizon, only which contributes to the imaginary part of the amplitude.
We can then perform the $r$-integration:
\begin{align}
\mbox{Im }S
&\simeq\mbox{Im}\int_{M}^{M-\omega}\int_{r_{in}}^{r_{out}}
\frac{2}{\pa_rf(r)\bigr|_{r=r_{+}}(r-r_{+})}dr dM' \nn
&=-2\pi\int_{M}^{M-\omega}\frac{dM^{\prime}}{\pa_rf(r)\bigr|_{r=r_{+}(M')}}.
\label{5-19}
\end{align}
For small $\omega$, this gives
\begin{align}
\mbox{Im }S&\simeq2\pi\omega\left[F(M)-\frac{\omega}{2}\frac{\pa}{\pa M}F(M)\right]\nn
&=\frac{2\pi\omega}{\pa_rf(r)\bigr|_{r=r_{+}\left(M\right)}}
-\frac{\pa}{\pa M}\left(\frac{1}{\pa_rf(r)\bigr|_{r=r_{+}\left(M\right)}}\right)\pi\omega^2,
\label{5-20}
\end{align}
where $F(M)\equiv\frac{1}{\pa_rf(r)\rvert_{r=r_{+}\left(M\right)}}$.
By using the result (\ref{5-20}), we obtain WKB probability amplitude
\begin{align}
\Gamma =e^{-2\;{\rm Im\,}S}\simeq e^{-\frac{2\pi}{\kappa}\omega
+2\frac{\pa}{\pa M}\left(\frac{1}{\pa_rf(r)\rvert_{r=r_{+}\left(M\right)}}\right)\pi\omega^2}
\label{5-21}
\end{align}
We thus see that the Hawking temperature is given precisely by the surface gravity
\bea
T_{H}=\frac{\pa_r f(r_+)}{4\pi}=\frac{\kappa}{2\pi},
\ena
and the quadratic term represents the correction by the backreaction of the radiation.

This is the general formula we get for the Hawking temperature and backreaction.
We find that the result for the Hawking temperature is precisely given by the surface
gravity, in agreement with the first approach.

\section{Discussions and Conclusions}
\label{Conclusion}

In this paper we have studied the Hawking radiation of a new class of black hole solutions
in the Einstein-Gauss-Bonnet theory. We first summarised the surprising result that
the mass and entropy of the black hole are zero, but it has nevertheless nonvanishing
Hawking temperature. The previous derivation of the Hawking temperature was
based on the Euclideanization of the geometry and it was not clear if this means
that the black hole really radiates or not. To check this, we have examined
the radiation by using the original Hawking method and find that it indeed radiates.
To confirm the result further, we took another derivation based on the tunneling approach.
We find that both method give consistent result that the black hole radiates,
and the Hawking temperature is given by the surface gravity.
However, it is not very clear in the second method until we compute the explicit
expression whether the temperature is related to the surface gravity.
Then slightly modifying the tunneling method, we have been able
to derive a general formula for the Hawking temperature which directly shows that
the result is related to the surface gravity.
Though some of the technical details may not be new, we believe that this result
is important and at least gives first step to understanding quantum aspect of this
kind of black holes.

Black hole solutions with a nonvanishing temperature and vanishing mass and entropy may
sound strange. However it is not so and there are many black holes having such
thermodynamical properties, for example, see~\cite{CLS,LLL}.
In the $R^2$ gravity, Lifshitz type solution is found, satisfying $1+2\a R=0$ where
$\a$ is the coefficient of $R^2$ term~\cite{CLS}. It has zero entropy but finite temperature.
The factor $1+2\a R$ plays the role of the effective coupling constant for polarization
graviton. The effective coupling is given by $G_{\rm eff}=G/(1+2\a R)$.
Thus the effective coupling diverges for the class of solutions with $1+2\a R=0$.
Wald's entropy is a quarter of the horizon area divided by the effective gravitational
constant~\cite{Wald}, and then the entropy vanishes.
This being so, one may expect that the fluctuations also vanish.
As a result, the entropy would vanish and then
the first law of the thermodynamics tells us that the mass should also vanish.
It was argued that this is also true here in \cite{CCO}.
As can be seen from \cite{MD}, the effective field equation for the four-dimensional
part is trivially satisfied because the coefficients in front of some gravitational
tensors are correspondent to the factor $1+2\a R$ discussed above for $R^2$ gravity
and vanish. In this sense, the effective coupling constants in the four-dimensional
viewpoint vanish identically. Thus the effective gravitational constant diverges
as in the $R^2$ gravity.
Still we find here that there is a Hawking radiation.
Once again we emphasize that the first law of thermodynamics $dM=T_H dS$ is satisfied
since the mass and entropy both vanish,
and there is no apparent inconsistency in the results.
Probably the mass of the black hole gets negative in the process of radiation.
The result may also appear to contradict the Clausius relation. If the black hole
emits Hawking radiation, there is a nonzero flow of heat from the black hole towards outside.
How this can be compatible with the vanishing entropy is left for future study.

One might suspect that we should consider the problem from the higher-dimensional point
of view. However what matters here is the four-dimensional radiation which is governed
by four-dimensional field equations in the black hole backgrounds. Our treatment should
be sufficient for such radiation.
There is another possibility that this might be a question to be asked for black holes
in the strong coupling regime.
Though our result indicates that we do have Hawking radiation, it may be
possible that the strong coupling effects modify the result.
At the moment, it is not clear to us if and how this could happen.
Obviously this system is worth studying further in connection with the quantum
fluctuations around the black holes, and is expected to shed light on the quantum
properties of black holes.

\section*{Acknowledgement}

We would like to thank R.-G. Cai and K. Umetsu for valuable discussions.
This work was supported in part by the Grant-in-Aid for
Scientific Research Fund of the JSPS (C) No. 20540283, No. 21$\cdot$09225 and
(A) No. 22244030.

%%%%%%%%%%%%%%%%%%%%%%%%%%%%%%%%%%%%%%%%%%


\begin{thebibliography}{99}
\bibitem{H}
  S.~W.~Hawking,
%  ``Particle Creation by Black Holes,''
  Commun.\ Math.\ Phys.\  {\bf 43} (1975) 199
  [Erratum-ibid.\  {\bf 46} (1976) 206].
\bibitem{PW}
  M.~K.~Parikh and F.~Wilczek,
%  ``Hawking radiation as tunneling,''
  Phys.\ Rev.\ Lett.\  {\bf 85} (2000) 5042
  [arXiv:hep-th/9907001].
\bibitem{RW}
  S.~P.~Robinson and F.~Wilczek,
  %``A Relationship between Hawking radiation and gravitational anomalies,''
  Phys.\ Rev.\ Lett.\  {\bf 95} (2005) 011303
  [arXiv:gr-qc/0502074];
  S.~Iso, H.~Umetsu and F.~Wilczek,
  %``Hawking radiation from charged black holes via gauge and gravitational
  %anomalies,''
  Phys.\ Rev.\ Lett.\  {\bf 96} (2006) 151302
  [arXiv:hep-th/0602146].
\bibitem{MD}
  H.~Maeda and N.~Dadhich,
%  ``Kaluza-Klein black hole with negatively curved extra dimensions in string
%  generated gravity models,''
  Phys.\ Rev.\  D {\bf 74} (2006) 021501
  [arXiv:hep-th/0605031];
%  H.~Maeda and N.~Dadhich,
%  ``Matter without matter: Novel Kaluza-Klein spacetime in
%  Einstein-Gauss-Bonnet gravity,''
  Phys.\ Rev.\  D {\bf 75} (2007) 044007
  [arXiv:hep-th/0611188].
\bibitem{CCO}
  R.~G.~Cai, L.~M.~Cao and N.~Ohta,
%  ``Black Holes without Mass and Entropy in Lovelock Gravity,''
  Phys.\ Rev.\  D {\bf 81} (2010) 024018
  [arXiv:0911.0245 [hep-th]].
\bibitem{GH}
  G.~W.~Gibbons and S.~W.~Hawking,
  %``Action Integrals and Partition Functions in Quantum Gravity,''
  Phys.\ Rev.\  D {\bf 15} (1977) 2752.
\bibitem{HL1}
  K.~Srinivasan and T.~Padmanabhan,
  %``Particle production and complex path analysis,''
  Phys.\ Rev.\  D {\bf 60} (1999) 024007
  [arXiv:gr-qc/9812028].
\bibitem{HL2}
  S.~Shankaranarayanan, T.~Padmanabhan and K.~Srinivasan,
  %``Hawking radiation in different coordinate settings: Complex paths
  %approach,''
  Class.\ Quant.\ Grav.\  {\bf 19} (2002) 2671
  [arXiv:gr-qc/0010042].
\bibitem{Va}
  E.~C.~Vagenas,
  %``Generalization of the KKW analysis for black hole radiation,''
  Phys.\ Lett.\  B {\bf 559} (2003) 65
  [arXiv:hep-th/0209185].
\bibitem{HL3}
  S.~Shankaranarayanan,
  %``Temperature and entropy of Schwarzschild-de Sitter space-time,''
  Phys.\ Rev.\  D {\bf 67} (2003) 084026
  [arXiv:gr-qc/0301090].
\bibitem{HL4}
  M.~Angheben, M.~Nadalini, L.~Vanzo, S.~Zerbini,
  %``Hawking radiation as tunneling for extremal and rotating black holes,''
  JHEP {\bf 0505 } (2005)  014.
  [hep-th/0503081].
\bibitem{HL5}
  M.~Arzano, A.~J.~M.~Medved, E.~C.~Vagenas,
  %``Hawking radiation as tunneling through the quantum horizon,''
  JHEP {\bf 0509 } (2005)  037.
  [hep-th/0505266].
\bibitem{HL6}
  A.~J.~M.~Medved, E.~C.~Vagenas,
  %``On Hawking radiation as tunneling with back-reaction,''
  Mod.\ Phys.\ Lett.\  {\bf A20 } (2005)  2449-2454.
  [gr-qc/0504113].
\bibitem{HL7}
  Q.~-Q.~Jiang, S.~-Q.~Wu,
  %``Hawking radiation of charged particles as tunneling from Reissner-Nordstrom-de Sitter black holes with a global monopole,''
  Phys.\ Lett.\  {\bf B635 } (2006)  151-155.
  [hep-th/0511123].
\bibitem{Ke}
  R.~Kerner and R.~B.~Mann,
  %``Tunnelling, Temperature and Taub-NUT Black Holes,''
  Phys.\ Rev.\  D {\bf 73} (2006) 104010
  [arXiv:gr-qc/0603019].
\bibitem{Ch}
  B.~D.~Chowdhury,
  %``Problems with Tunneling of Thin Shells from Black Holes,''
  Pramana {\bf 70} (2008) 593
  [arXiv:hep-th/0605197].
\bibitem{Ak}
  E.~T.~Akhmedov, V.~Akhmedova and D.~Singleton,
  %``Hawking temperature in the tunneling picture,''
  Phys.\ Lett.\  B {\bf 642} (2006) 124
  [arXiv:hep-th/0608098].
\bibitem{Hu}
  Y.~P.~Hu, J.~Y.~Zhang and Z.~Zhao,
  %``Massive particles-prime Hawking radiation via tunneling from the G.H.
  %dilaton black hole,''
  Mod.\ Phys.\ Lett.\  A {\bf 21} (2006) 2143
  [arXiv:gr-qc/0611026];
\bibitem{Mi}
  P.~Mitra,
  %``Hawking temperature from tunnelling formalism,''
  Phys.\ Lett.\  B {\bf 648} (2007) 240
  [arXiv:hep-th/0611265].
\bibitem{Wu}
  X.~n.~Wu and S.~Gao,
  %``Tunneling Effect Near Weakly Isolated Horizon,''
  Phys.\ Rev.\  D {\bf 75} (2007) 044027
  [arXiv:gr-qc/0702033].
\bibitem{LZ}
  C.~Z.~Liu and J.~Y.~Zhu,
  %``Hawking radiation and black hole entropy in a gravity ' $s$ rainbow,''
  Gen.\ Rel.\ Grav.\  {\bf 40} (2008) 1899
  [arXiv:gr-qc/0703055].
\bibitem{Pi}
  T.~Pilling,
  %``Tunneling derived from Black Hole Thermodynamics,''
  Phys.\ Lett.\  B {\bf 660} (2008) 402
  [arXiv:0709.1624 [gr-qc]].
\bibitem{Ki}
  S.~P.~Kim,
  %``Hawking Radiation as Quantum Tunneling in Rindler Coordinate,''
  JHEP {\bf 0711} (2007) 048
  [arXiv:0710.0915 [hep-th]].
\bibitem{SK}
  S.~Sarkar and D.~Kothawala,
%  ``Hawking radiation as tunneling for spherically symmetric black holes: A
%  Generalized treatment,''
  Phys.\ Lett.\  B {\bf 659} (2008) 683
  [arXiv:0709.4448 [gr-qc]].
\bibitem{LR}
  R.~Li and J.~R.~Ren,
  %``Dirac particles tunneling from BTZ black hole,''
  Phys.\ Lett.\  B {\bf 661} (2008) 370
  [arXiv:0802.3954 [gr-qc]].
\bibitem{CJYZ}
  D.~Y.~Chen, Q.~Q.~Jiang, S.~Z.~Yang and X.~T.~Zu,
  %``Fermions tunnelling from the charged dilatonic black holes,''
  Class.\ Quant.\ Grav.\  {\bf 25} (2008) 205022
  [arXiv:0803.3248 [hep-th]].
\bibitem{BM}
  R.~Banerjee and B.~R.~Majhi,
%  ``Quantum Tunneling Beyond Semiclassical Approximation,''
  JHEP {\bf 0806} (2008) 095
  [arXiv:0805.2220 [hep-th]];
%  ``Hawking black body spectrum from tunneling mechanism,''
  Phys.\ Lett.\  B {\bf 675} (2009) 243
  [arXiv:0903.0250 [hep-th]].
\bibitem{APS}
  E.~T.~Akhmedov, T.~Pilling and D.~Singleton,
  %``Subtleties in the quasi-classical calculation of Hawking radiation,''
  Int.\ J.\ Mod.\ Phys.\  D {\bf 17} (2008) 2453
  [arXiv:0805.2653 [gr-qc]].
\bibitem{ZCZ}
  B.~Zhang, Q.~y.~Cai and M.~s.~Zhan,
%  ``Hawking radiation as tunneling derived from Black Hole Thermodynamics
%  through the quantum horizon,''
  Phys.\ Lett.\  B {\bf 665} (2008) 260
  [arXiv:0806.2015 [hep-th]].
\bibitem{Mo}
  S.~K.~Modak,
  %``Corrected entropy of BTZ black hole in tunneling approach,''
  Phys.\ Lett.\  B {\bf 671} (2009) 167
  [arXiv:0807.0959 [hep-th]].
\bibitem{U}
  K.~Umetsu,
%  ``Hawking Radiation from Kerr-Newman Black Hole and Tunneling Mechanism,''
  Int.\ J.\ Mod.\ Phys.\  A {\bf 25} (2010) 4123
  [arXiv:0907.1420 [hep-th]];
%  K.~Umetsu,
%  ``Tunneling Mechanism in Kerr-Newman Black Hole and Dimensional Reduction
%  near the Horizon,''
  Phys.\ Lett.\  B {\bf 692} (2010) 61
  [arXiv:1007.1823 [hep-th]].
\bibitem{BMV}
  R.~Banerjee, B.~R.~Majhi and E.~C.~Vagenas,
  %``Quantum tunneling and black hole spectroscopy,''
  Phys.\ Lett.\  B {\bf 686} (2010) 279
  [arXiv:0907.4271 [hep-th]].
\bibitem{HL8}
  S.~H.~Mehdipour,
%  ``Hawking radiation as tunneling from a Vaidya black hole in noncommutative
%  gravity,''
  Phys.\ Rev.\  D {\bf 81} (2010) 124049
  [arXiv:1006.5215 [gr-qc]].
\bibitem{Ya}
  A.~Yale,
  %``Exact Hawking Radiation of Scalars, Fermions, and Bosons Using the Tunneling Method Without Back-Reaction,''
  Phys.\ Lett.\ B\ {\bf 697} (2011) 398
  [arXiv:1012.3165 [gr-qc]].
\bibitem{HL10}
  K.~Matsuno and K.~Umetsu,
  %``Hawking radiation as tunneling from squashed Kaluza-Klein black hole,''
  Phys.\ Rev.\  D {\bf 83} (2011) 064016
  [arXiv:1101.2091 [hep-th]].
\bibitem{VAD}
For a review, see
  L.~Vanzo, G.~Acquaviva, R.~Di Criscienzo,
  %``Tunnelling Methods and Hawking's radiation: achievements and prospects,''
  Class.\ Quant.\ Grav.\  {\bf 28 } (2011)  183001.
  [arXiv:1106.4153 [gr-qc]].
\bibitem{Wald}
  R.~M.~Wald,
  %``Black hole entropy is the Noether charge,''
  Phys.\ Rev.\ D\ {\bf 48} (1993) 3427
  [gr-qc/9307038].
\bibitem{CLS}
  R.~-G.~Cai, Y.~Liu and Y.~-W.~Sun,
  %``A Lifshitz Black Hole in Four Dimensional R**2 Gravity,''
  JHEP\ {\bf 0910} (2009) 080
  [arXiv:0909.2807 [hep-th]].
\bibitem{LLL}
  H.~Liu, H.~Lu and M.~Luo,
  %``On Black Hole Stability in Critical Gravities,''
  arXiv:1104.2623 [hep-th].
\end{thebibliography}
\end{document}